# Information Security Management in High Quality IS Journals:

# A Review and Research Agenda

**Sean B. Maynard**
School of Computing and Information Systems
The University of Melbourne
Parkville, Victoria, Australia
Email: seanbm@unimelb.edu.au

**Atif Ahmad**
School of Computing and Information Systems
The University of Melbourne
Parkville, Victoria, Australia
Email: atif@unimelb.edu.au

**Abstract**

*In the digital age, the protection of information resources is critical to the viability of organizations. Information Security Management (ISM) is a protective function that preserves the confidentiality, integrity and availability of information resources in organizations operating in a complex and evolving security threat landscape. This paper analyses ISM research themes, methods, and theories in high quality IS journals over a period of 30 years (up to the end of 2017). Although our review found that less than 1% of papers to be in the area of ISM, there has been a dramatic increase in the number of ISM publications as well as new emerging themes in the past decade. Further, past trends towards subjective-argumentative papers have reversed in favour of empirically validated research. Our analysis of research methods and approaches found ISM studies to be dominated by one-time surveys rather than case studies and action research. The findings suggest that although ISM research has improved its empirical backing over the years, it remains relatively disengaged from organisational practice.*

# Introduction

Organized, sophisticated and persistent threat actors have recently emerged as potent new risk for organizations (Ahmad et al., 2019;). These knowledgeable, well-trained and methodical human attackers can disrupt and destroy critical cyber infrastructures, deny organizations access to their own IT infrastructures and services (e.g. ransom-ware attacks), and steal sensitive information including Intellectual Property, trade secrets and customer data (Ahmad et al., 2019; Lemay et al., 2018). Such attacks may have negative consequences for organizations including: loss of competitive advantage, productivity, reputation and customer confidence; legal penalties and direct financial loss (Ahmad et al., 2021; Kotsias et al., 2022).

Effective Information Security Management (ISM) is critical to organizations if they are to survive in a complex and evolving security threat landscape. Drawing on various definitions of ISM in the extant literature, we define ISM as the structured application of formal, informal and technological controls with the objective of preserving the confidentiality, integrity and availability of information resources whilst maintaining strategic alignment with the organizational mission (Whitman & Mattord, 2014, Ashenden, 2008, Vermeulen & Von Solms, 2002, Alwi & Fan, 2009).

Organizations embarking on a program of ISM tend to draw on industry standards and associated white papers for guidance on how to meet their specific security requirements. In a series of seminal papers, Siponen (2006) and Siponen and Willison (2009) argue that industry standards have two key flaws: (1) they are too abstract and generic, focusing only on the existence of security processes and therefore lack specific guidance on *practice* that considers the different security requirements of organizations, and (2) the guidelines provided have not been empirically validated to the extent that they are a reliable source of advice to organizations.

Given the aforementioned flaws and the need for effective ISM, this paper aims to analyse the academic response in terms of the intellectual development of research in high quality Information Systems (IS) journals. Therefore, in this paper we ask: *To what extent does IS research contribute to ISM? What themes are studied? What approaches are used? What theories are adopted? What methodologies are used?* In this paper, we conduct an in-depth review of 'leading IS research' (i.e. publications listed in the AIS Senior Scholars' Basket of 8 Journals plus Information & Management and Decision Support Systems), with a view to analysing significant contributions to ISM.

This paper begins with a review of previous literature analyses in Information Security to (1) inform our research strategy, (2) inform our selection of classification frameworks, (3) motivate our analysis of the literature, and (4) structure the concepts in the review. Subsequently it describes the research methodology and then presents our detailed findings, discussion and research agenda. The paper concludes with key contributions and future research.

## Previous Literature Reviews of Information Security

There have been four substantive literature reviews in Information Systems Security in the past ten years (2007-2016) (Blake & Ayyagari, 2012; Silic & Back, 2014; Willison & Siponen, 2007; Zafar & Clark, 2009). Olijnyk (2015) did not conduct a literature review, but studied authors, citations and demographics in information security. The sections following compare and contrast the four substantive literature review's key characteristics.

### *Aims of the Review and Design of the Data Set*

All four literature reviews classify research in the discipline of Information Systems and provide an agenda for academics focused on increased understanding of trends, knowledge gaps, and an overall characterization of the intellectual development of the discipline area.

The authors of all the literature reviews argue that the rising trend of information security breaches in organizations calls for research into improved and effective Information Security. All reviews looked at 'Information Security' literature in general. They differed slightly in their criteria for the selection of papers (see "Design of Data Set" in Table 1), which lead to varying article data sets (the largest was 1,588 while the smallest was 137). The most widespread practice was to start from articles in the AIS Senior Scholars' Basket of Journals, then augment the article pool with additional security journals and/or conferences that publish Information Systems studies.

### *Classification Frameworks & Process*

The studies used either predefined classification frameworks to analyse the pool of literature or 'grounded theory' by constructing emergent frameworks derived from the literature itself. To engage in commentary on knowledge concentrations, knowledge gaps, trends in topics studied and intellectual development in general, the authors required a typology or framework of sub-topics that adequately represented the discipline of 'Information Security'. One avenue was to use the literature itself as a classification typology (Willison and Siponen, 2007), another was to use a predefined classification scheme (Zafar and Clark, 2009), a third was to adopt a hybrid method starting with the ISO 27002 industry standard augmented by a grounded theory approach (Silic and Back, 2014). Finally, an exception to the general trend was to use Latent Semantic Analysis – an objective method of analysis that measures term-to-term and document-to-document semantic similarities to infer relationships between documents (Blake and Ayyagari, 2012).

## *Outcomes & Limitations*

The primary outcomes of the reviews were the following: (1) the number of Information Security articles in leading Information Systems journals is low and has even decreased (Willison & Siponen, 2007); (2) Information Security research is underdeveloped in terms of theory as relatively few papers cite a theory and theories tend to be cited only once (Silic & Back, 2014; Willison & Siponen, 2007); (3) Information Security research suffers from a lack of empirical evidence as most papers are subjective-argumentative and relatively few papers presented empirical evidence (Silic & Back, 2014; Willison & Siponen, 2007); (5) Research is dominated by technical themes (Blake & Ayyagari, 2012; Zafar & Clark, 2009); (6) more experimentation is required to determine causal relations (Zafar & Clark, 2009); (7) there is a strong focus on particular security topics such as Privacy, Compliance, HR-related themes (Silic & Back, 2014); (7) there is a lack of research into specific areas such as Security Advisory Teams (Zafar & Clark, 2009), Business Continuity Management (Silic & Back, 2014) and social elements of security (Willison & Siponen, 2007).

*Table 1: Comparison of previous information security reviews*

|  | **Design of Data Set** | **Classification Process** | **Outcomes** |
|---|---|---|---|
| **Willison and Siponen (2007)** | **Data Set:** Papers from Information Systems & Information Security Journals over the last fifteen years. **Source:** Top 20 Information Systems journals and 3 Information Security journals **Size:** 1280 articles | **Framework(s): (1)** Laudan (1984) for theories, methods and aims; **(2)** Galliers (1992) for research classification; **(3)** 'grounded theory' for themes **Classification Process:** Reviewed article full text of, categorized themes using the classification framework; inter-rater reliability | **(1)** Overview of security research themes in Information Systems journals **(2)** Theories in security research in Information Systems journals **(3)** Research Methods in security research in Information Systems journals **(4)** Research Agenda |
| **Zafar and Clark (2009)** | **Data Set:** Papers from Information Systems Journals.<br>**Source:** (1) AIS Senior Scholars' Basket of Journals, and (2) the top 10% of journals listed in Peffers and Tang (2003)<br>**Size:** 137 articles | **Framework(s):** IBM Security Capability Reference Model<br>**Classification Process:** Reviewed each article (unclear if authors read the full text of papers) and categorized themes according to the framework. | **(1)** Synthesis of security themes in Information Systems journals<br>**(2)** Research Agenda |
| **Blake and Ayyagari (2012)** | **Data Set:** Papers from Information Systems & Information Security Journals.<br>**Source:** AIS Senior Scholars' Basket of Journals + 2 security journals<br>**Size:** 261 articles | **Framework(s):** None - applied factor analysis methods (Latent Semantic Analysis) to paper abstracts to determine the 'loading' of themes | **(1)** Security themes in Information Systems journals<br>**(2)** Demonstration of a repeatable method of analysing a body of research |
| **Silic and Back (2014)** | **Data Set:** Papers from Information Systems Journals.<br>**Source:** Determined by a 4 step process: (1) Journal search, (2) database search, (3) keyword search, (4) backward / forward search<br>**Size:** 1,588 articles | **Framework(s): (1)** Laudan (1984) for theories, methods & aims; **(2)** Galliers (1992) for research classification; **(3)** ISO 270002 extended by 'grounded theory' for themes<br>**Classification Process:** Reviewed full text of articles & categorized themes using the framework. No inter-rater reliability. | **(1)** Synthesis of security themes in Information Systems journals<br>**(2)** Theories used in security research in Information Systems journals<br>**(3)** Research Methods used in security research in Information Systems journals<br>**(4)** Research Agenda |

*Insights from the Literature Reviews*

Although our general research aims coincide with previous literature reviews, we have three further objectives. First, we are particularly interested in ISM research that informs organizational practice. This requires us to adopt a research methodology framework that comprehensively classifies the distinct types of practice-relevant research. Since this objective is not served by following the 'Latent Semantic Analysis' method adopted by Blake and Ayyagari (2012), we chose to follow the same method as Willison and Siponen (2007). Second, we are interested in determining which leading or 'high quality' journals are most suited to publishing research that informs ISM practice, therefore our data set is top tier journals only. Third, having a theme-based classification framework is necessary if one is to draw meaningful insights into what topic areas have been the focuses of research and what topics call for further research. However, topic classification frameworks are difficult to construct and highly subjective but simultaneously necessary in order to draw insightful analysis of literature. Thus, a grounded theory approach was used to analyse this literature so that a classification framework can be developed.

# Research Methodology

We used a five-stage grounded theory approach to produce a "theory-based, concept-centric" review of the literature (Wolfswinkel et al., 2011). Table 2 summarizes our approach.

*Define*

In line with this research approach we have already scoped the field of research (1.2 in Table 2) and appropriate sources (1.3 in Table 2) as our focus is on ISM research (as defined in the introduction) within high quality journals. We use the Association of Information Systems' (AIS) Senior Scholars' Basket of Journals. We also add Information & Management and Decision Support Systems, as these were recognised by many AIS special interest groups as also being highly regarded (Curry et al. 2016).

*Table 2: Five-Stage Grounded Theory Literature Review Method (Wolfswinkel et al. 2011)*

| **Tasks** | | |
|---|---|---|
| 1 Define | 1.1 | Define the criteria for inclusion/exclusion |
| | 1.2 | Identify the fields of research |
| | 1.3 | Determine the appropriate sources |
| | 1.4 | Decide on specific search terms |
| 2 Search | 2.1 | Search the literature |
| 3 Select | 3.1 | Refine the sample |
| 4 Analyse | 4.1 | Open coding |
| | 4.2 | Axial coding |
| | 4.3 | Selective coding |
| 5 Present | 5.1 | Represent and structure the content |
| | 5.2 | Structure the article |

In defining criteria for inclusion or exclusion (1.1 in Table 2) we selected any articles that focused on security within our sample set of journals, and included search terms (1.4 in Table 2) "security", "breach", "incident", "information assurance" etc. In addition to searching for these terms throughout articles we decided to manually review titles of articles that might have a security focus but did not use the identified terms.

*Table 3: Journals Used in this Review*

| Name | Journal Impact Factor ^ | 5 Year Impact Factor ^ | Start Year | End Volume | Total Papers Published | Total ISM Papers | % of ISM papers |
|---|---|---|---|---|---|---|---|
| Decision Support Systems | 3.847 | 4.903 | 1985 | 80 | 2740 | 16 | 0.58% |
| European Journal of Information Systems | 2.603 | 4.756 | 1991 | 24 | 848 | 17 | 2.00% |
| Information and Management | 4.120 | 5.346 | 1977 | 52 | 2015 | 26 | 1.29% |
| Information Systems Journal | 3.286 | 4.879 | 1991 | 25 | 485 | 7 | 1.44% |
| Information Systems Research | 2.457 | 6.069 | 1990 | 26 | 799 | 8 | 1.00% |
| Journal of Information Technology | 3.125 | 4.721 | 1986 | 30 | 747 | 1 | 0.13% |
| Journal of Management Information Systems | 3.013 | 4.560 | 1984 | 32 | 1141 | 14 | 1.23% |
| Journal of Strategic Information Systems | 4.000 | 4.851 | 1991 | 24 | 472 | 8 | 1.69% |
| Journal of the Association for Information Systems | 3.103 | 3.480 | 2000 | 16 | 436 | 6 | 1.38% |
| MIS Quarterly | 4.373 | 9.608 | 1977 | 40 | 1180 | 17 | 1.44% |
| **TOTAL** | | | | | **10863** | **120** | **1.10%** |
| ^ Values sourced from InCites Journal Citation Reports, 2018 on September 12, 2019. | | | | | | | |

### *Search*

In addition to conducting full text searches through all articles in the selected journals we conducted an article-by-article search manually. Each journal issue was reviewed, from its inception through to the end of 2017, inspecting the titles of papers to identify those that may potentially make significant contributions to ISM (this included all papers that were security orientated). During this process, detailed counts of the number of total papers published in each journal were recorded, as well as details of the identified ISM papers. The combination of our manual review and of our full text search ('2 Search' in Table 2) resulted in 313 candidate papers.

### *Select*

The researchers independently reviewed each of the candidate papers, reading the abstract, introduction, conclusion and skimming the results to identify those papers that focused on information security management. The researchers then met together to reconcile the results. They agreed that 116 papers were focusing on ISM, 193 were not ISM and 4 required further review. Subsequently, the researchers independently read the 4 in-doubt papers in full and came back for a second round of discussion. This resulted in complete agreement on the status of the last 4 papers. This process resulted in a total of 120 papers, which the researchers considered to be about information security management.

### *Analyse*

The researchers read each of the papers independently and text relevant to this study was highlighted. The researchers then used open coding (4.1 in Table 2) to identify the main concepts and insights of the articles (themes) within the framework of the study, focusing on the review criteria (see review criteria presented later). They then used an axial coding process (4.2 in Table 2) to identify the interactions between the themes identified in open coding. Finally, selective coding (4.3 in Table 2) was used to integrate and refine the identified themes, across each of the review criteria. During the coding process, we re-evaluated each of the articles to ensure that they were dealing with information security management. No further articles were discarded because of the coding process.

*Present*

The results and discussion parts of this paper presents the final stage of the Grounded Theory Literature Review Method (5 in Table 2).

## Review Criteria

*Themes*

Information security researchers have used various schemes in literature reviews to classify papers into themes (see Table 1). The classification schemes however consider the complete information security field, including a myriad of technological security aspects that are not relevant to ISM practices. Given these classifications are too abstract for a detailed analysis of the area of information security management, we used the grounded theory approach to identify the themes as evidenced from the papers (see Table 3).

*Table 3: Research Themes Identified*

| **Themes:** Incident Management, Security Policy Management, Security Risk Management, Security Education Training & Awareness, Technical Security Management, Intra-Organizational Liaison Management, Privacy, Security Compliance, Security Governance, Security Literature Analysis, Security Economics, Security Standards & Regulations, Security Culture / Behaviour, Security Methods, Security Requirements, Security Strategy |
|---|

*Research Methods*

We used Zhang and Li (2005) as our starting point as we found it to be the most comprehensive framework of research methods and identified the largest number and most relevant empirical methods (Table 4). We made two changes to this. First, we added an additional item to classify action research in order to distinguish it from general field studies (this was also done by Willison and Siponen, 2007). Second, we removed the distinction between interpretive and positivist case studies, as in Alavi and Carlson's (1992) original article. This was done because we were not concerned with how the case data was interpreted, but rather that the method was a case method.

*Table 4: Research Methods (descriptions can be found in Zhang and Li (2005)*

| | | |
|---|---|---|
| Non-Empirical | Conceptual Orientation | Defining Frameworks |
| | | Conceptual model of a process or structure |
| | | Conceptual overviews of ideas, theories, concepts |
| | | Theory from reference disciplines |
| | Illustration | Opinion (pure, or supported by examples) |
| | | Opinion (supported by personal experiences) |
| | | Description of a tool, technique, method, model |
| | Applied Concepts | Conceptual frameworks and applications |
| Empirical | Objects | Descriptions of types or classes of products, technologies, systems |
| | | Descriptions of a specific application, system, installation, program |
| | Events/process | Lab experiment |
| | | Field experiment |
| | | Field study |
| | | Action Research: |
| | | Case study |
| | | Survey |
| | | Development of instruments |
| | | Ex-post description of some project or event |
| | | Use of Secondary data |
| | | Interview |
| | | Delphi study |
| | | Focus group |

*Research Cycle*

The research cycle can be categorized using three stages: theory building, theory testing and theory refinement (Arnott & Pervan, 2005; Galliers, 1992; Neuman, 2000). In non-exploratory research, when the research is in the theory building stage, research questions are explored with the aim to eventually lead to a theory. Theory testing answers the research question through the design and execution of an appropriate research method. Theory refinement develops and tests the theory towards improvement. Shanks, Rouse, and Arnott (1993) argue that scholarship is the ability of the researcher to synthesize knowledge in their field, and from other fields, and apply this knowledge to the current problem domain. This can be evidenced by the research cycle, particularly in theory development and theory refinement. In this research, we gathered data on the particular stage of the research cycle that papers were focused on, i.e. *Theory Development*, *Theory Testing* and/or *Theory*

*Refinement*. Additionally, as we were interested in the impact of the research on practice, we also gathered data about whether the papers were about practice.

### *Level of Analysis*

The various research methods adopted by authors can be applied at different levels of analysis. We use four levels of analysis in information systems research: individual, group, organizational, and inter-organizational (societal) (proposed by Bariff and Ginzberg (1982)). From an ISM perspective, there are close parallels to information systems, especially as ISM is mainly about how people behave in relation to security (notwithstanding the technological aspects of information security).

### *Theories Used in Security Management Research*

To assess the extent to which theories have been used in ISM research, we draw on Laudan's (1984) 'reticulated model of science'. Laudan argues that a theory must make a distinct contribution to a study and must provide a good fit with research method. We endeavoured to comprehensively identify 'named' theories in ISM research to determine the extent to which theories inform the discourse in ISM research. Further, we introduce the research cycle classification– theory building, testing and refinement to better analyse the maturity of the theoretical contributions made to ISM.

### *Journal Contributions*

As one of our aims is to investigate a specific set of information systems journals, the collection of data on these journals is important. We wanted to assess their impact, which authors publish in them, what topics are published, what research methods are used are acceptable, and overall what percentage of papers are focused on Information Security Management. As such during the data collection we gathered additional statistics on journals such as their impact factors and publications volumes. This is in line with other literature reviews that report on similar statistics (Dibbern et al., 2001; Olijnyk, 2015; Zafar & Clark, 2009; Zhang & Li, 2005).

## Results

### *General Observations*

We found the number of ISM papers in IS journals to be quite low. Approximately one percent of papers published in our journal set were classified as being 'Information Security Management' (1.1 percent - 120 papers out of a total of 10863 papers). However, a more encouraging statistic can be drawn from Table 5 showing that 71% of all ISM publications were published in the last ten years.

*Table 5: ISM Publications by Journal over Time*

|  | Pre 1988 | 1988 - 1992 | 1993-1997 | 1998-2002 | 2003-2007 | 2008-2012 | 2013-2017 | Total |
|---|---|---|---|---|---|---|---|---|
| **DSS** |  |  |  |  |  | 5 | 11 | 16 |
| **EJIS** |  | 2 | 1 |  | 2 | 6 | 6 | 17 |
| **I&M** | 4 | 3 | 2 | 1 | 2 | 5 | 9 | 26 |
| **ISJ** |  |  |  | 1 | 1 | 1 | 4 | 7 |
| **ISR** |  | 1 |  |  | 1 | 4 | 2 | 8 |
| **JIT** |  |  |  |  |  |  | 1 | 1 |
| **JMIS** |  |  |  |  | 2 | 5 | 7 | 14 |
| **JSIS** |  |  |  | 1 | 3 | 2 | 2 | 8 |
| **JAIS** |  |  |  |  | 2 | 1 | 3 | 6 |
| **MISQ** |  | 3 | 1 | 1 | 1 | 7 | 4 | 17 |
| **Totals** | 4 | 9 | 4 | 4 | 14 | 36 | 49 | 120 |

DSS-Decision Support Systems, EJIS-European Journal of Information Systems, I&M-Information & Management, ISJ-Information Systems Journal, ISR-Information Systems Research, JIT-Journal of Information Technology, JMIS-Journal of Management Information Systems, JSIS-Journal of Strategic Information Systems, JAIS-Journal of the Association for Information Systems, MISQ-Management Information Systems Quarterly

Interestingly, almost 75% of all the articles were published in 5 out of the 10 journals, these being: I&M (22%), EJIS (14%), MISQ (14%), DSS (13%), JMIS (12%) (although MISQ had a 2010 special issue on ISM). Of the remaining journals, Journal of Information Technology has only ever published a single study.

### *Security Management Themes in IS Research*

Many of the papers identified (73%) studied more than one theme, which resulted in 244 research themes identified in the 120 papers (Table 6). The themes were concentrated on only 4 of the 16 themes in the typology (72% of paper themes). The most frequently studied topics were Information Security Policy Management (23%) followed by Security Culture/Behaviour (20%), Information Security Risk Management (18%) and finally Information Security Education, Training and Awareness (SETA) (12%). The remaining topics were observed less frequently in papers with the next highest being 5%. Interestingly, there was not a single study in Incident Management published in any of the surveyed journals. Other management topic areas such as Security Requirements, Privacy, Standards & Regulations, Governance and Intra-Organizational Communication have been studied less than 6 times. There have been only 2 analyses of security literature.

Table 6: Security Management Themes by Journal

| Themes/Journals | DSS | EJIS | I&M | ISJ | ISR | JIT | JMIS | JSIS | JAIS | MISQ | Totals | Theme % |
|---|---|---|---|---|---|---|---|---|---|---|---|---|
| Security Economics | 1 | | 1 | | 3 | | 3 | 1 | | 1 | 10 | 4% |
| Incident Management | | | | | | | | | | | 0 | 0% |
| Intra-Organisational Liaison Management | | 2 | | 1 | | | | | | | 3 | 1% |
| Privacy | | | | | | | | 1 | | 1 | 2 | 1% |
| Security Compliance | 2 | 1 | 2 | 1 | 2 | | 1 | 1 | 1 | 1 | 12 | 5% |
| Security Education Training & Awareness | 4 | 6 | 7 | 1 | 2 | | 1 | 1 | 2 | 5 | 29 | 12% |
| Security Governance | 1 | 1 | 1 | | 1 | | | | | 1 | 5 | 2% |
| Security Literature Analysis | | 1 | | 1 | | | | | | | 2 | 1% |
| Security Methods | 1 | 1 | 2 | | | | | | | | 4 | 2% |
| Security Policy Management | 3 | 10 | 10 | 5 | 4 | | 6 | 5 | 1 | 11 | 55 | 23% |
| Security Requirements | 1 | | | | | | 1 | | | | 2 | 1% |
| Security Risk Management | 7 | 2 | 10 | 2 | 4 | 1 | 7 | 2 | 1 | 7 | 43 | 18% |
| Security Standards & Regulations | 1 | 1 | | | | | | | 1 | 2 | 5 | 2% |
| Security Strategy | 2 | | 3 | | 1 | | 1 | 3 | | 1 | 11 | 5% |
| Security Culture / Behaviour | 7 | 8 | 8 | 4 | 2 | | 5 | 3 | 4 | 8 | 49 | 20% |
| Technological Security Management | 2 | | 5 | 1 | | | 4 | | | | 12 | 5% |
| Totals | 32 | 33 | 49 | 15 | 20 | 1 | 29 | 17 | 10 | 38 | 244 | |
| Journal % | 13% | 14% | 20% | 6% | 8% | 0% | 12% | 7% | 4% | 16% | | 100% |

DSS-Decision Support Systems, EJIS-European Journal of Information Systems, I&M-Information & Management, ISJ-Information Systems Journal, ISR-Information Systems Research, JIT-Journal of Information Technology, JMIS-Journal of Management Information Systems, JSIS-Journal of Strategic Information Systems, JAIS-Journal of the Association for Information Systems, MISQ-Management Information Systems Quarterly

None of the journals addressed all ISM themes. The most diverse journal was DSS (12 out of 16 themes). JIT, JAIS and ISJ were the least diverse journals with 1, 6 and 7 themes respectively. Some journals have a strong preference for certain themes. Information Security Policy Management has been studied 11 times in MISQ, 10 times in I&M, and 10 times in EJIS. Security Risk Management has been studied 10 times in I&M, however, four other journals also have a high proportion of studies in this area (7 studies in DSS, JMIS, MISQ).

Table 6 provides a useful overview of the themes of research publications categorized by journal; however, we can draw more in-depth insights by studying the distribution of such themes over time (see Table 7). As previously mentioned, the vast majority of ISM papers have appeared in the last ten years. As expected, some themes researched prior to 2006 continue to be studied with a similar regularity over the last ten years (Technological Security Management, Intra-Org Liaison Management & Security Governance). Also, unsurprisingly, some themes have become increasingly popular (Security Policy Management, Security Risk Management & Security Culture/Behaviour). However, interestingly the last decade has seen relatively new areas of research. These include Security Economics (from 1 in 2003-2007 to 4 in 2008-2012 & 5 in 2013-2017), Security Compliance (from 2 in 2003-2007, to 11

in 2013-2017), Security Requirements (3 in 2013-2017), and Security Standards and Regulations (from 1 in 2003-2007 to 2 in 2008-2013 and 2 in 2013-2017).

*Table 7: Themes over Time*

| Theme/Year | Pre 1988 | 1988 - 1992 | 1993- 1997 | 1998- 2002 | 2003- 2007 | 2008- 2012 | 2013- 2017 | Total |
|---|---|---|---|---|---|---|---|---|
| Security Economics | 0 | 0 | 0 | 0 | 1 | 4 | 5 | 10 |
| Incident Management | 0 | 0 | 0 | 0 | 0 | 0 | 0 | 0 |
| Intra-Organisational Liaison Management | 0 | 1 | 1 | 0 | 0 | 0 | 1 | 3 |
| Privacy | 0 | 1 | 0 | 0 | 1 | 0 | 0 | 2 |
| Security Compliance | 0 | 0 | 0 | 0 | 0 | 2 | 11 | 13 |
| Security Education Training & Awareness | 1 | 3 | 0 | 1 | 2 | 17 | 5 | 29 |
| Security Governance | 1 | 0 | 1 | 0 | 0 | 2 | 1 | 5 |
| Security Literature Analysis | 0 | 0 | 0 | 1 | 0 | 0 | 1 | 2 |
| Security Methods | 0 | 0 | 1 | 0 | 1 | 1 | 1 | 4 |
| Security Policy Management | 1 | 7 | 2 | 0 | 3 | 23 | 19 | 55 |
| Security Requirements | 0 | 0 | 0 | 0 | 0 | 0 | 3 | 3 |
| Security Risk Management | 2 | 5 | 1 | 3 | 6 | 12 | 14 | 43 |
| Security Standards & Regulations | 0 | 0 | 0 | 0 | 1 | 2 | 2 | 5 |
| Security Strategy | 0 | 1 | 0 | 1 | 1 | 3 | 4 | 10 |
| Security Culture / Behaviour | 1 | 2 | 1 | 0 | 2 | 21 | 23 | 50 |
| Technological Security Management | 2 | 1 | 1 | 1 | 2 | 2 | 3 | 12 |
| **TOTAL** | 8 | 21 | 8 | 7 | 20 | 89 | 93 | 246 |

### *Research Methods in ISM papers*

Some of the papers identified (17% or 20/120) used more than one research method, which resulted in 145 paper research methods identified in the 120 papers (Table 8).

*Table 8: Research Method in Journal*

| | | | DSS | EJIS | I&M | ISJ | ISR | JIT | JMIS | JSIS | JAIS | MISQ | Total |
|---|---|---|---|---|---|---|---|---|---|---|---|---|---|
| Non Empirical | Conceptual | Frameworks | | | 2 | | | | | | | | 2 |
| | | Theory from reference disciplines | 2 | 2 | | 1 | | | 1 | | 2 | 2 | 10 |
| | | Process / Structure | | | 1 | | | | | | | | 1 |
| | | Overview of Ideas | | 2 | 2 | 1 | | | 1 | | | | 6 |
| | Illustration | Pure Opinion - from examples | | 1 | 2 | | | | | | | 1 | 4 |
| | | Pure Opinion - from personal experience | | 1 | 4 | | | | | 2 | | | 7 |
| | | Description of Artifact | 2 | | 1 | | 1 | | 4 | | | | 8 |
| | Applied Concepts | Conceptual Frameworks / Applications | | | 4 | | | | 1 | 1 | 1 | | 7 |
| Empirical | Objects | Types or classes of Artifacts | | | | 1 | | | | | | | 1 |
| | | Specific Artifacts | 1 | | | | 1 | | | 1 | | | 3 |
| | Events / Processes | Lab Experiment / Simulation | 4 | 1 | | | | | 2 | 1 | 1 | 1 | 10 |
| | | Field Experiment | | | | | 2 | | | | 1 | | 3 |
| | | Field Study | | 1 | | | 1 | | | | | | 2 |
| | | Case Study | 1 | 3 | 1 | | | 1 | | 4 | | 1 | 11 |
| | | Survey | 6 | 6 | 13 | 4 | 4 | | 4 | 1 | 2 | 9 | 49 |
| | | Instrument Development | | | | | | | | | | | 0 |
| | | Ex-Post Description of Project / Event | | | | | | | | | | | 0 |
| | | Secondary Data | | | | | | | 2 | | | 2 | 4 |
| | | Interview | | | 4 | 1 | 1 | | | 1 | | 3 | 10 |
| | | Delphi Study | 1 | | 1 | | | | | | | | 2 |
| | | Focus Group | | | | | | | | | | 1 | 1 |
| | | Action Research | | 1 | | | | | | | | 3 | 4 |
| | | Totals | 17 | 18 | 35 | 8 | 10 | 1 | 15 | 11 | 7 | 23 | 145 |

DSS-Decision Support Systems, EJIS-European Journal of Information Systems, I&M-Information & Management, ISJ-Information Systems Journal, ISR-Information Systems Research, JIT-Journal of Information Technology, JMIS-Journal of Management Information Systems, JSIS-Journal of Strategic Information Systems, JAIS-Journal of the Association for Information Systems, MISQ-Management Information Systems Quarterly

The research methods used in ISM publications are dominated by empirical research (100 of the 145 methods used). Prior to 2000 there was relatively little empirical research published, however there was a dramatic increase in empirical research after this date (Table 9). The most widely used method is the empirical survey (49 times). The last decade, in particular, has seen a dramatic increase in the application of this method (from **3** in 2003 – 2007 to **18** in 2008-2012 and **20** in 2013 to 2017). Also, almost all journals have published an ISM paper using this method. Conceptual research methods have been used consistently over the years such that prior to 2000 ISM publications were dominated by such methods. Among these is the practice of applying theories drawn from outside the ISM discipline area (10 times). There has been an increase in this practice as well which is reflected in the increasing volume of papers published in ISM (**2** in 2001- 2005, **5** in 2006-2010, **5** in 2011-2015).

*Table 9: Research Method over Time*

|  |  |  | Pre 1988 | 1988-1992 | 1993-1997 | 1988-2002 | 2003-2007 | 2008-2012 | 2013-2017 | Total |
|---|---|---|---|---|---|---|---|---|---|---|
| Non Empirical | Conceptual | Frameworks |  | 1 |  |  | 1 |  |  | 2 |
|  |  | Theory from reference disciplines |  |  |  |  | 1 | 2 | 6 | 1 | 10 |
|  |  | Process / Structure | 1 |  |  |  |  |  |  | 1 |
|  |  | Overview of Ideas |  | 2 |  | 1 | 1 | 1 |  | 5 |
|  | Illustration | Pure Opinion - from examples | 1 | 1 |  |  |  | 1 | 1 | 4 |
|  |  | Pure Opinion - from personal experience | 3 |  |  | 1 | 1 | 1 | 1 | 7 |
|  |  | Description of Artifact |  |  |  |  |  | 3 | 1 | 4 | 8 |
|  | Applied Concepts | Conceptual Frameworks / Applications |  | 1 | 1 | 1 | 2 | 1 | 2 | 8 |
| Empirical | Objects | Types or classes of Artifacts |  |  |  |  |  | 1 |  | 1 |
|  |  | Specific Artifacts |  |  |  |  |  | 2 | 1 | 3 |
|  | Events / Processes | Lab Experiment / Simulation |  |  |  |  |  | 3 | 7 | 10 |
|  |  | Field Experiment |  |  |  |  |  | 1 | 2 | 3 |
|  |  | Field Study |  |  |  |  |  | 1 | 1 | 2 |
|  |  | Case Study |  |  | 1 |  | 3 | 3 | 4 | 11 |
|  |  | Survey |  | 5 | 2 | 1 | 3 | 18 | 20 | 49 |
|  |  | Instrument Development |  |  |  |  |  |  |  | 0 |
|  |  | Ex-Post Description of Project / Event |  |  |  |  |  |  |  | 0 |
|  |  | Secondary Data |  |  |  |  |  |  | 4 | 4 |
|  |  | Interview |  | 1 |  | 1 | 2 | 3 | 3 | 10 |
|  |  | Delphi Study |  |  |  |  |  |  | 2 | 2 |
|  |  | Focus Group |  |  |  |  |  | 1 |  | 1 |
|  |  | Action Research |  |  |  | 1 |  | 2 | 1 | 4 |
|  |  | Totals | 5 | 11 | 4 | 7 | 18 | 46 | 54 | 145 |

Given that ISM is a practice-based discipline, it is interesting to note that the number of times that researchers have opted to attach themselves to organizations or human subject experts is quite low - Case Study (11 times), Action Research (4 times), Focus Group (1 time), and Interviews (10 times) and mostly has occurred in the past 10 years. Some methods are rarely used in ISM publications. Among these are Delphi (2 times), conceptual frameworks (2 times), focus groups (1 time), field experiment (2 times), field study (2 times).

Both I&M and MISQ publish the most diverse set of research methods in ISM papers, with the latter more weighted towards empirical research rather than conceptual research. I&M, in particular, publishes survey-based research (13 of the 35 studies) whereas MISQ has published all of the three Action Research articles. All research methods that have been substantively used in the past are still actively used as reflected in the ISM articles published in the last five years.

Table 10 shows a breakdown of methods published in 5-year intervals from 1988. As pointed out earlier, the most striking observation is the steep increase in the number of surveys published in the last ten years (38 out of 49 times). From a practice perspective, there is good news in the significant reversal of the trend against practice-based research. The overwhelming majority of relevant research was published in the last ten years (case study, focus group & action research and interview methods used (17 out of 26 times).

With respect to building, testing and refining novel theoretical contributions in ISM, we found 99 out of 120 papers engaged in ISM theory-building, whereas 72 of the 99 proceeded to test ISM theory as well, and only 3 of these subsequently went on to refine ISM theory. Table 10 maps the three categories according to theme (note more than one theme exists per paper). Interestingly, the majority of theory building and testing occurred in Policy Management, Risk Management, and Economics whereas no theory development occurred in Technological Management, and Privacy.

*Table 10: Theme vs Research Cycle*

|  | Theory Building | Theory Building & Testing | Theory Refinement | Totals |
|---|---|---|---|---|
| **Security Economics** | 10 | 8 |  | 18 |
| **Incident Management** |  |  |  | 0 |
| **Intra-Organisational Liaison Management** | 2 | 2 |  | 4 |
| **Privacy** |  |  |  | 0 |
| **Security Compliance** | 11 | 9 | 1 | 21 |
| **Security Education Training & Awareness** | 27 | 23 | 1 | 51 |
| **Security Governance** | 4 | 4 |  | 8 |
| **Security Literature Analysis** | 2 |  |  | 2 |
| **Security Methods** | 2 | 1 | 1 | 4 |
| **Security Policy Management** | 47 | 38 | 2 | 87 |
| **Security Requirements** | 2 | 2 |  | 4 |
| **Security Risk Management** | 35 | 28 | 1 | 64 |
| **Security Standards & Regulations** | 4 | 2 |  | 6 |
| **Security Strategy** | 10 | 5 |  | 15 |
| **Security Culture / Behaviour** | 9 | 8 | 2 | 19 |
| **Technological Security Management** | 1 |  |  | 1 |
| **Totals** | 166 | 130 | 8 | 304 |

*Theories in ISM papers*

We found 58 individual theories cited in our data set of 120 papers. Table 11 shows theories used twice or more in ISM research. The number of papers that cited one or more theories was 61 (51%). The most common theories cited are Deterrence Theory (14 out of 63), Protection Motivation Theory (9 out of 55), Theory of Planned Behaviour (6 out of 55) and Agency Theory (4 out of 55). Game Theory and Rational Choice Theory have been used three times each. Each of the remaining 11 theories shown in Table 11 has been used twice. The other 41 theories identified in papers have only been used once.

*Table 11 Theory by Publication*

|  | DSS | EJIS | I&M | ISJ | ISR | JIT | JMIS | JSIS | JAIS | MISQ | Total |
|---|---|---|---|---|---|---|---|---|---|---|---|
| Deterrence theory |  | 4 | 2 | 1 | 2 |  | 1 | 1 |  | 3 | 14 |
| Protection motivation theory |  | 2 | 3 | 1 |  |  |  |  |  | 3 | 9 |
| Theory of planned behaviour |  | 2 | 2 |  |  |  |  |  | 1 | 1 | 6 |
| Agency theory | 4 |  |  |  |  |  |  |  |  |  | 4 |
| Dual Process Theory of Habituation | 1 |  |  |  |  |  | 2 |  |  |  | 3 |
| Game theory |  |  |  |  | 1 |  | 1 | 1 |  |  | 3 |
| Rational choice theory |  |  |  |  | 1 |  |  |  |  | 1 | 2 |
| Circuits of power | 1 |  |  |  |  |  |  |  |  | 2 | 2 |
| Control theory |  |  |  | 1 |  |  |  |  |  | 1 | 2 |
| Coping theory |  |  |  |  |  |  | 1 |  |  | 1 | 2 |
| Eye Movement Based Memory | 1 | 1 |  |  |  |  |  |  |  |  | 2 |
| Institutional theory |  |  | 2 |  |  |  |  |  |  |  | 2 |
| Reactance theory |  |  |  | 2 |  |  |  |  |  |  | 2 |
| Resource based theory |  |  | 1 |  |  |  | 1 |  |  |  | 2 |
| Social bond theory |  |  | 1 |  |  |  |  |  |  | 1 | 2 |
| Social cognitive theory |  |  | 1 |  |  |  |  |  |  | 1 | 2 |
| Structuration theory |  | 1 |  |  |  | 1 |  |  |  |  | 2 |

DSS-Decision Support Systems, EJIS-European Journal of Information Systems, I&M-Information & Management, ISJ-Information Systems Journal, ISR-Information Systems Research, JIT-Journal of Information Technology, JMIS-Journal of Management Information Systems, JSIS-Journal of Strategic Information Systems, JAIS-Journal of the Association for Information Systems, MISQ-Management Information Systems Quarterly

Authors tend to utilize theories when researching Security Economics, Security Compliance, SETA, Security Policy Management, Security Strategy, and Security Culture/Behaviour. Topics that tend not to cite theories include Privacy, Security Methods, Security Requirements, and Security Standards and Regulations.

## Discussion and Research Agenda

In response to the need for effective ISM this paper analyses ISM themes, theories, and methods published in leading IS journals. Our analyses led to a series of important findings.

Unsurprisingly, like Willison and Siponen (2007) and Silic and Back (2014) in the context of Information Security, we found the number of ISM research papers to be very low (~1%). The majority of papers were published in MISQ and I&M. Further, if we were to discount publications in special issues and outside the AIS Senior Scholars' Basket of 8, then the number of papers falls to 72.

Although these statistics are bad news for the ISM academic, there is however a silver lining. Most of the papers in ISM were published in the last ten years. In fact, the number of ISM publications increased by 157% from 2003-2007 to 2008-2012 and then by 36% from 2008-2012 to 2013-2017. This trajectory suggests that in the next 5 years the number of ISM publications in leading IS journals papers might approach 66 (assuming a further 36% growth) which makes the coming period a good time for academics to publish in ISM.

ISM research is concentrated on a few topics. Most papers (72%) focus on just 4 of the 16 themes: Policy Management, Risk Management, Culture/Behaviour, and Education, Training and Awareness. Interestingly, many research papers appear to be dominated by social rather than technological issues with a large number of behavioural theories playing a role in theory development culminating in a variance model that is ultimately tested using a survey-based approach. For ISM researchers, applying this research design to ISM topics such as Policy, Risk, Culture/Behaviour and SETA seems to be the most reliable way of getting published in the AIS Senior Scholars' Basket of Journals (plus Information and Management and Decision Support Systems). This finding points to a reversal of the trend seen in 2007 by Willison and Siponen (2007) which compelled them to point out that research into the social elements of security was lacking. This trend is however consistent with the Latent Semantic Analysis conducted Blake and Ayyagari (2012) which revealed 'Behavioural Aspects' and 'Security Design & Management' to be primary research topics in the Information Security discipline.

There are many fertile topic areas that are calling out for quality research. There has not been a single study in Incident Management[1]. This is surprising as a commonly cited justification for conducting research in ISM is the increasing number of incidents facing organizations. Similar arguments can be made for research into Security Requirements and Standards & Regulations – also areas of critical importance to organizations. Given the amount of attention paid to Strategy and Governance in the Management literature, it is unfortunate that researchers have not drawn on the Management discipline to inform Strategy and Governance in the ISM context. We believe that the lack of research in these areas cannot be remedied without assistance from Journal editors. For example, more special issues on ISM research that explicitly encourage research in these areas would go a long way to addressing these gaps.

These findings are relatively consistent with some of the findings of previous literature reviews. We agree with Zafar and Clark (2009) that governance research is lacking (although this is not specific to Security Advisory Teams) and we also agree with Silic and Back (2014) that Business Continuity Management requires more focus, specifically incident management. We did not however find a substantive focus on Privacy and Compliance related topics as was reported by Silic and Back (2014), this might be owing to their larger data set and broader focus on Information Security which would significantly change the relative proportions of topics studied against a discipline area.

---

[1] At the time this paper was published in arxiv at least one paper on incident response appeared in a Basket of 8 journal - Kotsias et al., 2022

Most ISM papers (60%) tend to develop their own theory - this typically involves both building and testing. Although many papers use existing theories to inform their own theory development, they tend not to 'close the loop' by refining these existing theories themselves. Further, we could not find any ISM papers dedicated to testing and refining ISM theories (i.e. without first building their own theory). The findings also suggest that high quality IS journals do on occasion accept ISM papers that engage only in theory building without theory testing. On the one hand, this finding is rather surprising given most IS journals are very much focused on theory contributions, on the other hand the finding is good news for ISM academics as it suggests theory testing is not always a condition for acceptance in IS journals.

In general, theoretical development in ISM papers remains poor. Only 60% of ISM papers cited one or more theories. Most theory-informed research is focused on a few topics whereas other topics tended not to attract any theory references at all. Theories were frequently used to explain human behaviour when researching Security Compliance, SETA, Security Policy Management, Security Strategy, and Security Culture/Behaviour. Theories tend to be cited only once and authors tend not to revisit the theories or build on previous theory development in papers (as previously mentioned). This finding is consistent with both Willison and Siponen (2007) and Silic and Back (2014).

However, we found a dramatic increase in the proportion of papers published with empirical evidence. In their literature review, Willison and Siponen (2007) concluded that the Information Security discipline was intellectually 'hamstrung' because of the dominance of subjective-argumentative research and the lack of empirical research. This conclusion certainly applied to ISM at the time. However, in the last ten years, empirical research has dramatically increased with the surge of survey-based studies (see Table 9). This is a promising platform upon which ISM researchers can build.

We believe a critical challenge facing ISM research is the lack of engagement with organizational practice. We find particularly troubling the fact that much empirical research uses the survey method (where researchers are detached and disengaged from organizations) rather than case study and action research. Why is this the case? This is an interesting question as it has significant implications for ISM researchers as well as journal editors and organizations.

The most popular answer is that organizations are reluctant to give security researchers the access they need in the organization (Kotulic & Clark, 2004). However, this possibility can be discounted as there are plenty of case studies published in second tier journals such as Computers & Security (C&S) and the International Journal of Information Management (IJIM); which suggests that such research is certainly occurring and security authors are getting access to organizations. For example, a number of ISM case studies have been published in the last five years (e.g. Dhillon, Syed, and Pedron (2016) & da Veiga and Martins (2015) on Security Culture; Brender and Markov (2013) & Ozkan and Karabacak (2010) on Risk Management; Ahmad, Hadjkiss, and Ruighaver (2012), Ahmad, Maynard, and Shanks (2015) and Ahmad et al., 2020 on Incident Management; and Kolkowska and Dhillon (2013) on Policy Management). Another possible answer is that authors are not submitting such papers to journals because the long turnaround times render the practice-based data obsolete. For example, using the article history statistics of I&M made available post-2009, we determined the average turnaround time of I&M articles to be 2 years and 2 months. The longest review times were 64 months and 48 months whereas the shortest times were 13 and 15 months. EJIS did not fare much better with the average being 19 months, the maximum being 45 months and the minimum 10 months. Such turnaround times are widely known to be quite typical of this subset of journals.

The third possibility is that journals themselves are rejecting practice-based research possibly because (1) the sample sizes are too small to generalize (relative to sample sizes in surveys), (2) journals are simply uninterested in practice-based research and are more focused on theory, (3) it is difficult to find reviewers that can engage with and understand / assess organizational security practices to the extent that they are comfortable with accepting the paper. These arguments may explain why authors find the survey method more appealing. Authors can test theory-based variance models by collecting large amounts of empirical data without the need to access the internal practices of organizations and generalize from the findings.

The potential ramifications on industry from the lack of empirically validated and practice-relevant research are quite significant. If we are to accept that there is a community of professionals wanting to learn about the practice of security management then they are unable to get quality guidance from the academic community in a timely manner. In fact, the field of ISM suffers because key management practice areas such as incident management, risk management, and strategy are not subjected to studies that give insight into practice and enable positive change to occur. ISM academics are being implicitly encouraged to prefer quantitative survey-based research approaches over qualitative, exploratory methods.

Where should academics submit their ISM papers? Based on the proportion of ISM papers published in respective journals, the best candidates are MISQ and I&M. I&M may not be an option for many academics as it is not in the Senior Scholars' Basket of Journals, however it is an option for others (for example for Australian academics I&M is ranked in the top tier (A* on ACPHIS (2013) / ABDC (2013)). Unfortunately, like most of the journals used in this research, I&M has relatively long lead times for review as mentioned before. The long lead times in reviewing and the high risk of rejection (papers may be rejected after years have passed in the review process) make it particularly difficult for early career researchers to build a track record to be competitive for tenured appointments.

For ISM academics employed by institutions that adopt a more flexible interpretation of 'quality', there are options outside the top tier Information Systems journals. For example, the International Journal of Information Management (5y Impact Factor of 6.327), and Computers & Security (5y Impact Factor of 3.476) routinely publish ISM articles. Further, their rapid turnaround times (typically in the order of 2 to 3 months) make them extremely attractive alternatives especially for academics trying to build a track record of quality publications in a short amount of time.

A key contribution of this paper is our emergent taxonomy of ISM themes or practice areas (see Table 12). We identified the themes using a grounded theory approach when codifying research papers and then reconciled our classification of themes against those appearing in other review papers (see Table 3). Our taxonomy provides the most comprehensive, discriminating and parsimonious list of ISM areas compared to the other taxonomies and is an indispensable tool for researchers to review ISM literature, identify gaps in ISM knowledge, and organize ISM themes and ideas (Reisman, 2006).

*Table 12: Research Themes Classification*

| Theme | Source |
|---|---|
| Incident Management | Alshaikh et al. (2014); Silic and Back (2014) |
| Security Policy Management | Alshaikh et al. (2014); Silic and Back (2014); Willison and Siponen (2007); Zafar and Clark (2009); Maynard et al. (2011) |
| Security Risk Management | Alshaikh et al. (2014); Blake and Ayyagari (2012); Silic and Back (2014); Willison and Siponen (2007); Zafar and Clark (2009) |
| Security Education Training & Awareness | Alshaikh et al. (2014); Silic and Back (2014); Willison and Siponen (2007) |
| Technological Security Management | Alshaikh et al. (2014); Silic and Back (2014) |
| Intra-Organizational Liaison Management | Alshaikh et al. (2014); Silic and Back (2014) |

| Privacy | Blake and Ayyagari (2012); Zafar and Clark (2009) |
|---|---|
| Security Compliance | Silic and Back (2014); Zafar and Clark (2009) |
| Security Governance | Blake and Ayyagari (2012); Willison and Siponen (2007); Zafar and Clark (2009) |
| Security Literature Analysis | Willison and Siponen (2007) |
| Security Economics | Blake and Ayyagari (2012); Willison and Siponen (2007); Zafar and Clark (2009) |
| Security Standards & Regulations | Silic and Back (2014); Willison and Siponen (2007) |
| Security Culture / Behaviour | Willison and Siponen (2007) |
| Security Methods | Emergent from the literature |
| Security Requirements | Emergent from the literature |
| Security Strategy | Emergent from the literature |

## Conclusions

This paper reports on the extent to which ISM research is published in high quality IS journals. Our findings indicate that although approximately one percent of all IS publications focused on ISM themes, the number of publications has been dramatically increasing over the last ten years. Further, we note previous trends towards subjective-argumentative papers have reversed in favour of empirically validated research. However, the overwhelming majority of these studies utilize survey methods rather than case study and action research – suggesting ISM academics are not engaging in practice-driven research to the extent that they are attached and internal to the organization. A key outcome of our study is an emergent taxonomy of ISM themes based on a grounded theory analysis of literature. The taxonomy is a comprehensive, discriminating and parsimonious list of ISM areas and is an indispensable tool for researchers to review ISM literature, identify gaps in ISM knowledge, and organize ISM themes and ideas.

There are a number of avenues of future work that follow from this research. We focused our analysis of ISM papers on the AIS Senior Scholars' Basket of Journals (plus Information and Management and Decision Support Systems). Future studies may consider broadening the size and scope of the data set. This can be done in many different ways such as focusing on second tier IS journals, focusing on security journals, or adopting a keyword search strategy using "Information Security Management" etc. Conferences can be targeted in the same way. Future work can use a different set of classification frameworks in research method, theory, and theme to provide additional insights. For example, a more granular ISM taxonomy in each theme area can be used to analyse the contributions of ISM publications leading to a more detailed research agenda for that ISM theme area. Practice-based classification frameworks such as that of Engaged Scholarship (see Mathiassen and Nielsen (2008)) can be used to assess the level of engagement of ISM research in organizations. ISM academics can benefit from such analysis in terms of building a practice-based research agenda.

# References

ABDC. (2013). ABDC Journal Quality List - Australian Business Deans Council. Retrieved from http://www.abdc.edu.au/pages/abdc-journal-quality-list-2013.html

ACPHIS. (2013). IS Journal Ranking - Australian Council of Professors and Heads of Information Systems. Retrieved from http://www.acphis.org.au/index.php/is-journal-ranking

Ahmad, A., Maynard, S.B., Desouza, K.C., Kotsias, J., Whitty, M., & Baskerville, R.L., (2021). How can Organizations Develop Situation Awareness for Incident Response? A Case Study of Management Practice. *Computers & Security*. Vol 101. (pp. 1-15).

Ahmad, A., Desouza, K. C., Maynard, S. B., Naseer, H., & Baskerville, R. L. (2020). How Integration of Cyber Security Management and Incident Response Enables Organizational Learning. *Journal of the Association for Information Science and Technology*, 71(8), 939-953.

Ahmad, A., Desouza, K. C., Maynard, S. B., Naseer, H., & Baskerville, R. L. (2020). How Integration of Cyber Security Management and Incident Response Enables Organizational Learning. *Journal of the Association for Information Science and Technology*, 71(8), 939-953.

Ahmad, A., Hadjkiss, J., & Ruighaver, A. B. (2012). Incident Response Teams - Challenges in Supporting the Organizational Security Function. *Computers & Security, 31*(5), 643-652.

Ahmad, A., Maynard, S. B., & Shanks, G. (2015). A Case Analysis of Information Systems and Security Incident Responses. *International Journal of Information Management, 35*(6), 717-723. doi:10.1016/j.ijinfomgt.2015.08.001

Alavi, M., & Carlson, P. (1992). A Review of MIS Research and Disciplinary Development. *Journal of Management Information Systems, 8*, 45-62.

Alshaikh, M., Ahmad, A., Maynard, S. B., & Chang, S. (2014). *Towards a Taxonomy of Information Security Management Practices in Organisations*. Paper presented at the 25th Australasian Conference on Information Systems, Auckland, New Zealand.

Alwi, N. H. M., & Fan, I. S. (2009, November). Information security management in e-learning. In *2009 International Conference for Internet Technology and Secured Transactions,(ICITST)* (pp. 1-6). IEEE.

Arnott, D., & Pervan, G. (2005). A critical analysis of decision support systems research. *Journal of Information Technology, 20*(2), 67-87.

Ashenden, D. (2008). Information Security management: A human challenge?. *Information security technical report*, 13(4), 195-201.

Bariff, M. L., & Ginzberg, M. J. (1982). MIS and the behavioral sciences: research patterns and prescriptions. *ACM SIGMIS Database, 14*(1), 19-26.

Blake, R., & Ayyagari, R. (2012). Analyzing information systems security research to find key topics, trends, and opportunities. *Journal of Information Privacy and Security, 8*(3), 37-67.

Brender, N., & Markov, I. (2013). Risk perception and risk management in cloud computing: Results from a case study of Swiss companies. *International Journal of Information Management, 33*(5).

Clark, J. G., Au, Y. A., Walz, D. B., & Warren, J. (2011). Assessing researcher publication productivity in the leading information systems journals: A 2005-2009 update. *Communications of the Association for Information Systems, 29*(1), 459-504.

Curry, W.L.; Dennis, A.R.; Nickerson, R.; Niederman, F.; Vogel D. (2016) "Interim Report on the Senior Scholars Journal Basket Review", Thirty Seventh International Conference on Information Systems, Dublin. http://aisel.aisnet.org/cgi/viewcontent.cgi?article=1374&context=icis2016, Accessed 11/9/17.

da Veiga, A., & Martins, N. (2015). Improving the information security culture through monitoring and implementation actions illustrated through a case study. *Computers & Security, 49*, 162-176.

Dhillon, G., & Backhouse, J. (2001). Current directions in IS security research: towards socio-organizational perspectives. *Information Systems Journal, 11*(2), 127-153.

Dhillon, G., Syed, R., & Pedron, C. (2016). Interpreting information security culture: An organizational transformation case study. *Computers & Security, 56*, 63-69.

Dibbern, J., Goles, T., Hirschheim, R., & Jayatilaka, B. (2001). Information systems outsourcing: a survey and analysis of the literature. *ACM SIGMIS Database, 35*(4), 6-102.

Galliers, R. D. (1992). Choosing Information Systems Research Approaches. In R. D. Galliers (Ed.), *Information Systems Research: Issues, Methods and Practical Guidelines* (pp. 144-163).

IBM. (2009). IBM Security Reference Model. Retrieved from http://www-935.ibm.com/services/au/gbs/bus/html/IntroducingIBMFramework.pdf

Kolkowska, E., & Dhillon, G. (2013). Organizational power and information security rule compliance. *Computers & Security, 33*, 3-11.

Kotsias, J., Ahmad, A., & Scheepers, R. (2022). Adopting and Integrating Cyber-Threat Intelligence in a Commercial Organisation, *European Journal of Information Systems*, pp 1-17.

Kotulic, A. G., & Clark, J. G. (2004). Why There Aren't More Information Security Research Studies. *Information and Management, 41*, 597-607.

Laudan, L. (1984). *Science and Values*. CA, USA.: Berkeley: University of California Press.

Mathiassen, L., & Nielsen, P. A. (2008). Engaged scholarship in IS research. *Scandinavian Journal of Information Systems, 20*(2).

Maynard, S.B., Ruighaver, A.B. & Ahmad, A. (2011, Dec), Stakeholders in Security Policy Development. Paper presented at the 9th Information Security Management Conference, Perth, Australia. (pp. 182-188). Edith Cowan University.

Neuman, W. (2000). *Social Research Methods: Qualitative and Quantitative Approaches*: Allyn and Bacon.

Olijnyk, N. V. (2015). A quantitative examination of the intellectual profile and evolution of information security from 1965 to 2015 *Scientometrics, 105*(2), 883-904.

Oliveira, T., & Martins, M., F. (2011). Literature Review of Information Technology Adoption Models at Firm Level. *The Electronic Journal Information Systems Evaluation Volume, 14*(1), 110-121.

Ozkan, S., & Karabacak, B. (2010). Collaborative risk method for information security management practices: A case context within Turkey. *International Journal of Information Management, 30*(6).

Peffers, K., & Tang, Y. (2003). Identifying and Evaluating the Universe of Outlets for Information Systems Research: Ranking the Journals. *The Journal of Information Technology Theory and Application, 5*(1), 63-84.

Pervan, G. P. (1998). A review of research in Group Support Systems: leaders, approaches and directions. *Decision Support Systems, 23*, 149-159.

Reisman, A. (2006). A taxonomic view of illegal transfer of technologies: A case study. *Journal of Engineering and Technology Management*, *23*(4), 292-312.

Shanks, G. G., Rouse, A., & Arnott, D. (1993). *A Review of Approaches to Research and Scholarship in Information Systems* (3/93). Retrieved from

Sidorova, A., Evangelopoulos, N., Valacich, J. S., & Ramakrishnan, T. (2008). Uncovering the intellectual core of the information systems discipline. *MIS Quarterly, 23*(3), 467-482.

Silic, M., & Back, A. (2014). Information security: Critical review and future directions for research. *Information management & computer security, 22*(3), 279-308.

Siponen, M. (2006). Information security standards focus on the existence of process, not its content. *Communications of the ACM, 49*(8), 97-100.

Siponen, M. T., & Willison, R. (2009). Information Security Management Standards: Problems and Solutions. *Information and Management, 46*, 267-270.

Soomro, Z. A., Shah, M. H., & Ahmed, J. (2016). Information security management needs more holistic approach: A literature review. *International Journal of Information Management, 36*(2), 215-225.

Vermeulen, C., & Von Solms, R. (2002). The information security management toolbox–taking the pain out of security management. *Information management & computer security*, *10*(3), 119-125.

Vermolen, R. (2010). *Reflecting on is business model research: Current gaps and future directions*. Paper presented at the Proceedings of the 13th Twente Student Conference on IT, University of Twente, Enschede, Netherlands.

Vessey, I., V., R., & R.L., G. (2002). Research in information systems: An empirical study of diversity in the discipline and its journals. *Journal of Management Information Systems, 19*(2), 129-174.

Whitman, M. E., & Mattord, H. J. (2014). *Principles of Information Security*: Course Technology, Cengage Learning.

Willison, R., & Siponen, M. (2007). *A critical assessment of IS security research between 1990-2004*. Paper presented at the Proceedings of 15th European Conference on ISs, St. Gallen, Switzerland.

Wolfswinkel, J. F., Furtmueller, E., & Wilderom, C. P. (2013). Using grounded theory as a method for rigorously reviewing literature. *European journal of information systems*, *22*(1), 45-55.

Zafar, H., & Clark, J. G. (2009). Current state of information security research in IS. *Communications of the Association for Information Systems, 24*(1), 34.

Zhang, P., & Li, N. (2005). The intellectual advancement of human-computer interaction research: A critical assessment of the MIS literature (1990-2008). *Journal of the Association for Information Systems, 6*(11), 227-292.